\documentclass[12pt,intlim,righttag]{article}

\usepackage{amsmath,amsthm}
\usepackage{amssymb}
\usepackage{amsfonts}

\newcommand{\essinf}{\operatornamewithlimits{ess\,inf}}

\newcommand\beq{\begin{equation}}
\newcommand\eeq{\end{equation}}
\newcommand{\ld}{\lambda}
\newcommand{\vp}{\varphi}

\newcommand{\td}{\tilde}

\newcommand{\lbl}{\label}

\newcommand{\lb}{\lambda}

\newcommand{\la}{\langle}
\newcommand{\ra}{\rangle}

\newcommand{\wt}{\widetilde}
\newcommand{\wh}{\widehat}
\newcommand{\be}{\begin}
\newcommand{\ee}{\end}

\newcommand{\sg}{\sigma}

\theoremstyle{Theorem}

\newtheorem{prop}{\ \ \ Proposition}[section]

\theoremstyle{corollary}

\theoremstyle{remark}

\theoremstyle{definition}

\newtheorem{rem}{\ \ \ Remark}

\numberwithin{equation}{section}

\begin{document}

\title{$L^2$-approximating pricing under restricted information}

\author{M. Mania $^{1),2)}$ , R. Tevzadze $^{1),3)}$ and T. Toronjadze $^{1),2)}$}

\date{~}
\maketitle

\begin{center}
$^{1)}$ Georgian--American University, Business School,
3, Alleyway II,\\ Chavchavadze Ave. 17, a, Tbilisi, Georgia,

E-mail: toronj333@yahoo.com  \\[2mm]
$^{2)}$ A. Razmadze Mathematical Institute, 1, M. Aleksidze St.,
Tbilisi, Georgia, \\E-mail: mania@rmi.acnet.ge \\[2mm]
$^{3)}$Institute of Cybernetics, 5, S. Euli St., Tbilisi, Georgia,\\
E-mail:rtevzadze@posta.ge
\end{center}

\begin{abstract}

We consider the mean-variance hedging problem under partial
information in the case where the flow of observable events
does not contain the full information on the underlying asset price
process. We introduce a martingale equation of a new type and characterize
the optimal strategy in terms of the  solution of this equation.
We give relations between this equation and  backward stochastic differential
equations for the value process of the problem.

\bigskip



\end{abstract}

\section{Introduction}

We assume that the dynamics of the price process of the asset
traded on a  market is described by a continuous semimartingale
$S=(S_t,t\in[0,T])$ defined on a filtered probability space
$(\Omega,{F}, {\cal F}=({\cal F}_t,t\in[0,T],P)$, satisfying the
usual conditions, where ${F} ={\cal F}_T$ and $T <\infty$ is the
fixed time horizon. Suppose that the interest rate is equal to
zero and the asset price process satisfies the structure
condition, i.e.,  the process $S$ admits the decomposition
\begin{equation}\label{str22}
S_t=S_0+M_t+\int_0^t\lambda_ud\la M\ra_u,\;\;\;\la\lambda\cdot M\ra_T<\infty\;\;\;a.s.,
\end{equation}
where $M$ is a continuous ${\cal F}-$local martingale and
$\lambda$ is a $\cal F$-predictable  process.

Let us introduce an additional filtration smaller than $\cal F$
$$
{\cal G}_t\subseteq {\cal F}_t,\;\;\;\;\text{for every}\;\;\; t\in
[0,T].
$$
The filtration $\cal G$ represents the information that the hedger
has at his disposal, i.e., hedging strategies have to be
constructed using only information available in $\cal G$.

Let $H$ be a $P$-square integrable ${\cal F}_T$-measurable random
variable, representing the payoff of a contingent claim at time
$T$.

 We consider the mean-variance hedging problem
\begin{equation}\label{mvh}
\text{to minimize}\;\;\;\;\; E[
(X^{x,\pi}_T-H)^2]\;\;\;\;\text{over all}\;\;\;\;\pi\in\Pi({\cal
G}),
\end{equation}
where $\Pi({\cal G})$ is a class of $\cal G$-predictable
$S$-integrable processes. Here
 $X^{x,\pi}_t=x+\int_0^t\pi_udS_u$ is the wealth process starting
from initial capital $x$, determined by the self-financing trading
strategy $\pi\in\Pi({\cal G})$.

In the case $\cal G=\cal F$ of complete information the
mean-variance hedging problem was introduced by F\"ollmer and
Sondermann \cite{F-S} in the case when $S$ is a martingale and
then developed by several authors for price process admitting a
trend (see, e.g., \cite{D-R}, \cite{H}, \cite{S92},\cite{S94},
\cite{S}, \cite{G-L-Ph}, \cite{H-P-Sc}).

Asset pricing with partial information under various setups has
been considered. The mean-variance hedging problem under partial
information was first studied by Di Masi, Platen and Runggaldier
(1995) when the stock price process is a martingale and the prices
are observed only at discrete time moments. For a general
filtrations and when the asset price process is a martingale this
problem was solved by Schweizer (1994) in terms of $\cal
G$-predictable projections. Pham (2001) considered the
mean-variance hedging problem  for a general semimartingale model,
assuming that the observable filtration contains the augmented
filtration ${\cal F}^S$ generated by the asset price process $S$
\begin{equation}\label{fsg}
 {\cal F}^S_t\subseteq {\cal G}_t ,\;\;\;\;\text{for every}\;\;\; t\in [0,T].
\end{equation}
In this paper, using the variance-optimal martingale measure with
respect to the filtration $\cal G$ and suitable Kunita-Watanabe
decomposition, the theory developed by Gourieroux, Laurent and
Pham (1998) and Rheinl\"ander and Schweizer (1997) to the case of
partial information was extended.

If ${\cal F}^S_t\subseteq {\cal G}_t$, then the price process is a
$\cal G$-semimartingale and the sharp bracket $\la M\ra$ is $\cal
G$-adapted.
If $\cal G$ is not containing ${\cal F}^S$, then $S$ is not a
$\cal G$-semimartingale and the problem is more involved. We focus our attention to the case
when the filtration $\cal G$ of observable events does not contain the full information about the
asset price process $S$
and  solve the problem (\ref{mvh}) in this case under following additional assumptions:

A) $\la M\ra$ and $\lambda$ are $\cal G$-predictable,

B) any $\cal G$- martingale is a $\cal F$-local martingale,

C) the filtration $\cal G$ is continuous, i.e., all $\cal G$-local
martingales are continuous,

D) there exists a martingale measure for $S$ that satisfies the Reverse
H\"older condition.

Denote by $\widehat{Y}_t$  the process
$E(Y_t|{\cal G}_t)$- the $\cal G$-optional projection of $Y$.
Condition A) implies that
$$
\widehat S_t=E(S_t|{\cal G}_t)=S_0+\int_0^t\lambda_ud\la
M\ra_u+\widehat M_t.
$$

Let
\begin{equation*}
H_t= EH+\int_0^th_udM_u+L_t
\end{equation*}
and
\begin{equation*}
H_t= EH+\int_0^th^{\cal G}_ud\widehat M_u+L^{\cal G}_t
\end{equation*}
be the Galtchouk-Kunita-Watanabe (GKW) decompositions of
$H_t=E(H|{\cal F}_t)$ with respect to local martingales $M$ and
$\widehat M$, where $h, h^{\cal G}$ are $\cal F$-predictable
process and $L, L^{\cal G}$ are local martingales strongly
orthogonal to $M$ and $\widehat M$ respectively. We shall use also
notations:
$$\rho^2_t=\frac{d\la\widehat M\ra_t}{d\langle M\rangle_t},
\;\; {\widetilde h}_t= \widehat{h^{\cal G}_t}\rho^2_t-{\widehat
h}_t \;\;\;\; \text{and}\;\;\wt H=\widehat H_T-\int_0^T\frac{\wt
h_t}{1-\rho^2_t}d\widehat S_t.
$$

We introduce the following martingale equation
\begin{align}\lbl{pl}
\widetilde Y_T=\wt
H-\int_0^T\frac{1}{1-\rho_t^2}\left[\ld_t\widetilde Y_t
+\rho_t^2\widetilde\psi_t \right]\left(\ld_td\la M\ra_t+d\widehat
M_t\right).
\end{align}
The solution of this equation is a pair $(\wt Y,\wt\psi)$, where
$\wt Y$ is a square integrable martingale and $\wt\psi$ is defined by
the GKW decomposition of $\tilde Y$
$$
\wt Y_t=\tilde Y_0+\int_0^t\td\psi_u\wh M_u+L_t,\;\;\;\;\la\wh M,L\ra=0.
$$
Now we formulate the main result of the paper which is proved in section 3.

{\bf Theorem.} Let conditions A)-D) be satisfied. Assume also that
$E\td H^2 <\infty$ and $\rho^2_t< 1$ for all $t\in[0, T]$. Then there exists a unique
solution $(\tilde Y,\tilde\psi)$ of equation (\ref{pl})
 and the strategy $\pi^*$ is optimal if and only if it admits the representation
\begin{equation}\label{op5}
\pi^*_t=\frac{1}{1-\rho_t^2}\left(\widetilde h_t
+\ld_t\widetilde Y_t+{\rho_t^2\widetilde\psi_t}\right).
\end{equation}

In section 4 (see propositions \ref{pro1.5} and  \ref{pro1.4}) we
establish the connection between equation (\ref{pl}) and BSDEs for the value
process of the problem (\ref{mvh}) derived in
\cite{mtt}.  It was shown in  \cite{mtt} that the optimal strategy is determined
by
\begin{equation}\label{trpl31}
    \pi_t^*  =\frac{\lb_tV_t^{H}+\rho_t^2\vp_t^H-\wh X_t^{\pi^*}(\lb_tV_t+\rho_t^2\vp_t)}
            {1-\rho_t^2+\rho_t^2V_t},\;\;\;\wh X_0^{\pi^*}=x,
\end{equation}
where the tryples $(V, \varphi, L)$ and $(V^H,\vp^H, L^H)$  satisfy the following system of BSDEs
\begin{equation}\label{trpl32}
    dV_t  =\frac{(\lb_tV_t+\rho_t^2\vp_t)^2}{1-\rho_t^2+\rho_t^2V_t}\,d\langle M\rangle_t +
        \vp_t\,d\wh{M}_t+dL_t, \quad V_T=1,
 \end{equation}
\begin{equation}\label{trpl33}
    dV_t^{H}=\frac{(\lb_tV_t+\rho_t^2\vp_t)(\lb_tV_t^{H}+\rho_t^2\vp_t^H)}
            {1-\rho_t^2+\rho_t^2V_t}\,d\langle M\rangle_t
     + \vp_t^Hd\wh{M}_t+dL_t^H, \;\;\;V_T^{H}=\wt H.
\end{equation}
Here $L$ and $L^H$ are ${\cal G}$-local martingales strongly orthogonal to $\wh M$.

Note that, to construct the optimal strategy by (\ref{op5}) we need to solve only
equation (\ref{pl}), which is easier to solve than equation (\ref{trpl32}), whereas for the construction
of the optimal strategy by  (\ref{trpl31})   one should solve equation (\ref{trpl32}) and two linear
equations (\ref{trpl31}) and (\ref{trpl33}).
On the other hand
the construction by (\ref{pl}), (\ref{op5})  does not contain  the
case of the full information (since in this case $\rho^2=1$ and the integral in (\ref{pl}) is not defined),
but the construction (\ref{trpl31})- (\ref{trpl33}) includes this case directly.

The relations between these equations are as follows:

If $(\tilde Y,\tilde\psi)$ is a solution of (\ref{pl}) for $H$ equal to strictly positive constant $c$,
then the processes $\tilde Y_t$,
$c-\int_0^t\pi_u^*d\wh S_u$ are strictly positive and  the process
$$
V_t=\frac{\td Y_t}{ c-\int_0^t\pi_u^*d\wh S_u},
$$
where $\pi^*$ is defined by (\ref{op5}), satisfies the BSDE   (\ref{trpl32}).

On the other hand, if the tryples $(V, \varphi, L)$ and $(V^H,\vp^H, L^H)$  satisfy (\ref{trpl32})-(\ref{trpl33}),
then the  pair $(\wt Y, \wt\psi)$, where $\wt Y_t=V_t^{H}-\wh X_t^{\pi^*}V_t$ and
$\wt\psi_t=\vp_t^H-V_t\pi^*_t-\vp_t\wh X_t^{\pi^*}$ ($\pi^*$ and $\wh X_t^{\pi^*}$ are defined by (\ref{trpl31})),
is a solution  of equation (\ref{pl}).

 In section 5 we consider a diffusion market model which
consists of two assets $S$ and $\eta$, where $S_t$ is a state of a
process being controlled and $\eta_t$ is the observation process.
Suppose that $S_t$ and $\eta_t$ are governed by
$$
dS_t=\mu_tdt+\sigma_tdw^0_t,
$$
$$
d\eta_t=a_tdt+b_tdw_t,
$$
where $w^0$ and $w$ are Brownian motions with correlation $\rho$
and the coefficients $\mu, \sigma, a$ and $b$ are ${\cal
F}^\eta$-adapted. So, in this case ${\cal F}_t={\cal
F}_t^{S,\eta}$ and the flow of observable events is ${\cal
G}_t={\cal F}^\eta_t$. We give in the case of markovian
coefficients solution of the problem (\ref{mvh}) in terms of
parabolic differential equations (PDE) and an explicit solution
when coefficients  and the contingent claim are constants.

\

\section{Main definitions and auxiliary facts}

\

Denote by ${\mathcal M}^e({\cal F})$ the set of equivalent
martingale measures for $S$, i.e., set of probability measures $Q$
equivalent to $P$ such that $S$ is a  $\cal F$-local martingale
under $Q$.

Let
$$
{\mathcal M}^e_2({\cal F})=\{Q\in{\mathcal M}^e({\cal F}):
EZ_T^2(Q)<\infty\},
$$
where $Z_t(Q)$ is the density process (with respect to the
filtration $\cal F$) of $Q$ relative to $P$.

{\bf Remark 2.1.}  Since $S$ is continuous, the
 existence of an equivalent martingale measure and the Girsanov
 theorem imply that the structure condition (\ref{str22}) is satisfied.

 Note that the density process $Z_t(Q)$
of any element $Q$ of ${\cal M}^e({\cal F})$ is expressed as an
exponential martingale of the form
$$
{\cal E}_t(-\lambda\cdot M+N),
$$
where $N$ is a $\cal F$-local martingale strongly orthogonal to
$M$ and ${\cal E}_t(X)$ is the Doleans-Dade exponential of $X$.

If the local martingale $Z^{min}_t={\cal E}_t(-\lambda\cdot M)$ is
a true martingale, $dQ^{min}/dP= Z^{min}_TdP$ defines
an equivalent probability measure called the minimal martingale
measure for $S$.

Recall that a measure $Q$
satisfies the Reverse H\"older inequality $R_2(P)$ if there exists
a constant $C$ such that
$$
E\big(\frac{Z_T^2(Q)}{Z_\tau^2(Q)}|{\cal F}_{\tau}\big)\le C,
\:\:\:\:\:P-a.s.
$$
for every $\cal F$-stopping time $\tau$.

{\bf Remark 2.2.} If there exists a measure $Q\in{\cal M}^e({\cal
F})$ that satisfies the Reverse H\"older inequality $R_2(P)$, then
according to Kazamaki \cite{Kz} the martingale $M^Q=-\lambda\cdot
M+N$ belongs to the class $BMO$ and hence $-\lambda\cdot M$ also
belongs to $BMO$, i.e.,
\begin{equation}\label{bmo}
E\big(\int_\tau^T\lambda^2_ud\langle M\rangle_u|{\cal
F}_\tau\big)\le const
\end{equation}
for every stopping time $\tau$. Therefore, it follows from Kazamaki \cite{Kz}
that ${\mathcal E}_t(-\lambda\cdot M)$ is
a true martingale. So, condition D) implies that the minimal martingale measure
exists (but $Z^{min}$ is not necessarily square integrable).

For all unexplained notations concerning the martingale theory used below we
refer the reader to \cite{D-M},\cite{L-Sh2},\cite{J}.

Let $\Pi({\cal F})$ be the space of all $\cal F$-predictable
$S$-integrable processes $\pi$ such that the stochastic integral
$$
(\pi\cdot S)_t=\int_0^t\pi_udS_u,\;\; t\in[0,T],
$$
is in the ${\cal S}^2$ space of semimartingales , i.e.,
$$
E\big(\int_0^T\pi^2_sd\langle M\rangle_s\big)+
E\big(\int_0^T|\pi_s\ld_s|d\langle M\rangle_s\big)^2<\infty.
$$
Denote by $\Pi({\cal G})$ the subspace of $\Pi({\cal F})$ of $\cal
G$-predictable strategies.

{\bf Remark 2.3.} Since $\lambda\cdot M\in BMO$ (see Remark 2.2),
it follows from the proof of Theorem 2.5 of Kazamaki \cite{Kz}
$$
E\big(\int_0^T|\pi_u\lambda_u|d\langle M\rangle_u\big)^2=
E\la|\pi|\cdot M,|\lambda|\cdot M\ra_T^2
$$
$$
\le 2||\lambda\cdot M||_{\rm BMO} E\int_0^T\pi_u^2d\langle
M\rangle_u<\infty.
$$
Therefore,  under condition D) the strategy $\pi$ belongs to the
class $\Pi({\cal G})$ if and only if $E\int_0^T\pi^2_sd\langle
M\rangle_s<\infty$.

Let us make a comment on condition B).

{\bf Remark 2.4.} Condition $B)$ is satisfied if and only if the
$\sigma$-algebras ${\cal F}_t$ and ${\cal G}_T$ are conditionally
independent given ${\cal G}_t$ for all $t\in[0,T]$ (see Theorem
9.29 from Jacod 1978).
Note that  one can weaken this condition imposing that any ${\cal G}$-martingale is a
$\wt G$- local martingale, where $\wt G$ is the augmented filtration generated by ${\cal F}^S$ and
${\cal G}$. This condition is satisfied if
${\cal F}_t^S\subseteq {\cal G}_t$. In this case instead of (\ref{str22}) one should use the
decomposition
\begin{equation}\label{ino}
S_t=S_0+\int_0^tE(\lambda_u|\wt G_u)d\langle M\rangle_u + N_t,
\end{equation}
where
\begin{equation}\label{inom}
N_t=M_t+ \int_0^t[\lambda_u- E(\lambda_u|{\wt G}_u)]d\langle N\rangle_u
\end{equation}
is a ${\wt G}$-local martingale.

Define $J_T^2(\cal F)$ and $J_T^2(\cal G)$ as  spaces of terminal
values of stochastic integrals, i.e.,
$$
J_T^2({\cal F})=\{(\pi\cdot S)_T:\pi\in\Pi({\cal F})\}.
$$
$$
J_T^2({\cal G})=\{(\pi\cdot S)_T:\pi\in\Pi({\cal G})\}.
$$

Now we recall some known assertions from the filtering theory.

Let $A=(A_t,t\in[0,T])$ be a RCLL process and there is a sequence
$(\tau_n, n\ge1)$ of $\cal G$-stopping times such that
$E\int_0^{\tau_n}|dA_u|<\infty$ for all $n\ge1$. Then there exists
a unique $\cal G$-predictable process $A^p$ of finite variation
(see Jacod 1978), called a $\cal G$-dual projection of $A$ such
that
$$
E(A_t|{\cal G}_t)- A^p_t\;\;\;\;\text{is a}\;\;{\cal
G}-\text{local martingale}.
$$

Throughout the paper we use the notation
$\widehat\Phi_t=E[\Phi_t|{\cal G}_t]$ for  any process $\Phi$.

For convenience, we give the proof of the following  assertion, which
is proved similarly to \cite{L-Sh2}.

\be{prop}\lbl{p2.2} If conditions $A), B)$ and $C)$ are satisfied,
then for any $\cal F$-local martingale $M$ and any $\cal G$-local
martingale $m^{\cal G}$
\begin{equation}\label{fg}
\widehat M_t=E(M_t|{\cal G}_t)= \int_0^tE\big(\frac{d\la M,m^{\cal
G}\ra_u}{d\la m^{\cal G}\ra_u}|{\cal G}_u\big)dm^{\cal G}_u
+L^{\cal G}_t,
\end{equation}
where $L^{\cal G}$ is a local martingale orthogonal to $m^{\cal
G}$. \ee{prop}

{\it Proof.} Since $\wh M$ is a continuous $\cal G$-local martingale, it admits
the GKW decomposition
\begin{equation}\label{kfg}
\widehat M_t=E(M_t|{\cal G}_t)= \int_0^tf_udm^{\cal G}_u
+L^{\cal G}_t,
\end{equation}
where $f_u=\frac{d\la\wh M,m^{\cal G}\ra_u}{d\la m^{\cal G}\ra_u}$. Thus, it is sufficient to show that
$d\la m^{\cal G}\ra_tdP$ -a.e.
\begin{equation}\label{kfg1}
\frac{d\la\wh M,m^{\cal G}\ra_u}{d\la m^{\cal G}\ra_u}=
 E\big(\frac{d\la M,m^{\cal
G}\ra_u}{d\la m^{\cal G}\ra_u}|{\cal G}_u\big).
\end{equation}
Since  $m^{\cal G}$ is  $F$-local martingale (by condition B), the process $M_tm^{\cal G}_t-\la M, m^{\cal G}\ra_t$ is also
a $\cal F$-local martingale. Conditions A) and C) imply that $M_tm^{\cal G}_t$ and $\la M, m^{\cal G}\ra_t$ are
$\cal G$-locally integrable. Therefore the processes   $E(M_tm^{\cal G}_t-\la M, m^{\cal G}\ra_t|{\cal G}_t)$   and
$E(\la M, m^{\cal G}\ra_t|{\cal G}_t)-\la M, m^{\cal G}\ra^p_t$ are $\cal G$-local martingales and hence the process
\begin{equation}\label{kfg2}
E(M_tm^{\cal G}_t|{\cal G}_t)-\la M, m^{\cal G}\ra^p_t
\end{equation}
is also a $\cal G$- local martingale.

On the other hand   $E(M_tm^{\cal G}_t|{\cal G}_t)=\wh M_t m^{\cal G}_t$  and the process
 $\wh M_t m^{\cal G}_t-\la\wh M, m^{\cal G}\ra_t$ is a $\cal G$-local martingale. Therefore
 the process
$$
E(M_tm^{\cal G}_t|{\cal G}_t)-\la\wh M, m^{\cal G}\ra_t
$$
is also a ${\cal G}$- local martingale. This, together with (\ref{kfg2}), implies that
\begin{equation}\label{kfg3}
\la\wh M, m^{\cal G}\ra_t=  \la M, m^{\cal G}\ra^p_t.
\end{equation}
But
$$
 \langle M, m^{\cal G}\rangle^{p}=
 \big(\int_0^t\frac{d\la M,m^{\cal G}\ra_u}{d\la m^{\cal
G}\ra_u}d\la m^{\cal G}\ra_u\big)^p=\int_0^tE\big(\frac{d\la M,m^{\cal
G}\ra_u}{d\la m\ra_u}|{\cal G}_u\big)d\la m^{\cal G}\ra_u,
$$
which proves equality  (\ref{kfg1}) and (\ref{fg}) holds.\qed
\be{cor} For any $\cal G$-predictable $S$-integrable process $\pi$
\begin{equation}\lbl{flttt}
\widehat{(\pi\cdot S)_t}=E\big(\int_0^t\pi_udS_u|{\cal G}_t)=
\int_0^t\pi_ud\widehat S_u.
\end{equation}
\ee{cor}
 {\it Proof}. It follows  from  proposition 2.1 that for any
$\cal G$-predictable, $M$-integrable process $\pi$ and any $\cal
G$-martingale $m^{\cal G}$ that
$$
\la\widehat{(\pi\cdot M)},m^{\cal G}\ra=
\int_0^t\pi_uE\big(\frac{d\la M,m^{\cal G}\ra_u}{d\la m^{\cal
G}\ra_u}|{\cal G}_u\big)d\la m^{\cal G}\ra_u
$$
$$
=\int_0^t\pi_ud\la\widehat M,m^{\cal G}\ra_u=\la\pi\cdot\widehat
M,m^{\cal G}\ra_t.
$$
Hence, for any $\cal G$-predictable, $M$-integrable process $\pi$
\begin{equation}\label{mg}
\widehat{(\pi\cdot M)_t}=E\big(\int_0^t\pi_sdM_s|{\cal
G}_t)=\int_0^t\pi_sd\widehat M_s.
\end{equation}
Since $\pi, \lambda$ and $\langle M\rangle$ are $\cal
G$-predictable, from (\ref{mg}) we obtain
(\ref{flttt})

{\bf Remark 2.5.}
In particular, equality (\ref{kfg3}) implies that
\begin{equation}\label{duw}
\langle M, \widehat M\rangle^{p}=\langle\widehat M\rangle
\end{equation}
and
\begin{equation}\label{dul}
\langle M, L\rangle^p=0
\end{equation}
if $L$ is a $\cal G$-local martingale orthogonal to $\widehat M$.

\be{lem}\lbl{lem2} Let conditions {\rm A)--C)} be satisfied and
$\widehat M_t=E(M_t|{\cal G}_t)$. Then $ \la \widehat M\ra $ is
absolutely continuous w.r.t $ \la M\ra $ and
$$
\rho^2_t=\frac{d \la \widehat M\ra _t}{d \la M\ra _t}\le 1.
$$
Moreover if $A=\{(\omega,t): \rho_t^2=1\}$ then  a.s. for all $t$
\begin{equation}\label{rho1}
\int_0^tI_{A}(u)dM_u=\int_0^tI_A(u)d\widehat M_u.
\end{equation}
\ee{lem}

{\it Proof.} By (\ref{mg}) for any bounded ${\cal G}$-predictable
process $f$
\begin{align}\lbl{l4}
\notag & E\int_0^tf_s^2d \la \widehat M\ra _s
\\
\notag &= E\left(\int_0^t f_sd\widehat M_s\right)^2
= E\left(E\left(\int_0^t f_sdM_s\big|{\cal G}_t\right)\right)^2\\
& \le E\left(\int_0^t f_sdM_s\right)^2\le E\int_0^t f_s^2d \langle
M\rangle_s
\end{align}
which implies that $ \la \widehat M\ra $ is absolutely continuous
w.r.t $ \langle M\rangle $, i.e.,
$$
 \la \widehat M\ra _t=\int_0^t\rho^2_sd \langle M\rangle _s
$$
for a ${\cal G}$-predictable process $\rho$. Moreover (\ref{l4})
implies that the process $\la M\ra - \la\wh M\ra$ is increasing
and hence $\rho^2\le 1$ $\mu^{\langle M\rangle}$ a.e.

Let us show now the equality (\ref{rho1}). It is evident that
$\int_0^tI_A(u)d\la M\ra_u= \int_0^tI_A(u)d\la M\ra_u$. Since the
set $A$ is $\cal G$-predictable and $\langle M, \widehat
M\rangle^{p}=\langle\widehat M\rangle$ by Proposition 2.2
$$
E\big(\int_0^tI_{A}(u)dM_u-\int_0^tI_A(u)d\widehat M_u\big)^2=
E\int_0^tI_A(u)d\la M-\widehat M\ra_u
$$
$$
=E\int_0^tI_A(u)d\la M-\widehat M\ra_u^{\cal
G}=E\int_0^tI_A(u)d\la M\ra_u- E\int_0^tI_A(u)d\la M\ra_u=0.
$$
 \qed
\be{cor} If $\rho_t^2=1$ for all $t$, then $M=\wh M$ and therefore
$M$ is a $\cal G$- local martingale . \ee{cor}
 We shall use the following Lemma proved in \cite{D-S}.
\be{lem}\lbl{polk} Let $N=(N_t,t\in[0,T])$ be a square integrable
martingale such that $N_0>0$. Let $\tau=\inf\{t:N_t\le0\}\wedge T$
\footnote{$)$\,It is assumed that $\inf\emptyset=\infty$ and
$a\wedge b$ denotes $\min\{a,b\}$ }$)$ be a predictable stopping
time announced by a sequence of stopping times $(\tau_n;n\ge1).$
Then
$$
E\left(\frac{N_T^2}{N_{\tau_n}^2}\big|{\cal
G}_{\tau_n}\right)\to\infty\;\text{on the set}\; (N_\tau=0)
$$
\ee{lem}
{\it Proof.}
\begin{align}\lbl{levy}
1=E\left(\frac{N_T}{N_{\tau_n}}\big|{\cal G}_{\tau_n}\right)
=E\left(\frac{N_T}{N_{\tau_n}}I_{(N_\tau=0)}\big|{\cal G}_{\tau_n}\right)\\
\le E^{\frac{1}{2}}\left(\frac{N_T^2}{N_{\tau_n}^2}\big|{\cal
G}_{\tau_n}\right) E^{\frac{1}{2}}\left(I_{(N_\tau=0)}|{\cal
G}_{\tau_n}\right)
\end{align}
By the Levy theorem $\lim_{n\to\infty}E\left(I_{(N_\tau=0)}|{\cal
G}_{\tau_n}\right)=I_{(N_\tau=0)}$ is equal to 0 on the set
$(N_\tau=0)$. Therefore it follows from (\ref{levy}) that
$E\left(\frac{N_T^2}{N_{\tau_n}^2}\big|{\cal G}_{\tau_n}\right)\to
\infty$ on $(N_\tau=0)$.\qed

\

\section{Mean-variance hedging and Forward-Backward \\equation}

 Let $X^*=X^{0,\pi^*}$ be the wealth process  corresponding to the optimal strategy
$\pi^*$.  Let
$H_t=E[H|{\cal F}_t]$ and let
\begin{equation}\label{hhh2}
H_t=E(H|{\cal F}_t)=  EH+\int_0^th_udM_u+L_t
\end{equation}
be the Galtchouk-Kunita-Watanabe (GKW) decomposition of $H_t$,
where $L$ is a martingale orthogonal to $M$ and $h$ is $\cal
F$-predictable $M$-integrable process. We shall use also the GKW
decomposition of $H_t=E(H|{\cal F}_t)$ with respect to the local
martingale $\widehat M$
\begin{equation}\label{htg}
H_t= EH+\int_0^th^{\cal G}_ud\wh{M}_u+L^{\cal G}_t.
\end{equation}
Here $h^{\cal G}$ is a $\cal F$-predictable process and $L^{\cal
G}$ is a $\cal F$- local martingale strongly orthogonal to
$\widehat M$.

It follows from  Proposition \ref{p2.2} ( applied for $m^{\cal
G}=\widehat M$) and  Lemma \ref{lem2} that

\begin{equation}\label{hmbar}
\la E(H|{\cal G}_.),\widehat M\ra_t=\int_0^tE(h^{\cal G}_u|{\cal
G}_u)d\la\widehat M\ra_u= \int_0^t\widehat{h_u^{\cal
G}}\rho^2_ud\la M\ra_u.
\end{equation}

\be{prop}\label{pr1} Assume that $\la M\ra$ is $\cal
G$-predictable.Then the optimal strategy $\pi^*$ of optimization
problem $(\ref{mvh})$ satisfies the relation
\begin{equation}\lbl{str}
\pi^*_t=E\left(h_t+\psi_t+\ld_tH_t+\ld_tY_t-\ld_tX^*_t|{\cal
G}_t\right),
\end{equation}
where the triple $(Y,\psi,L),\;\la L,M\ra=0$ is a solution of BSDE
\begin{equation}\lbl{bsde}
dY_t=\pi^*_t\ld_td\la M\ra_t+\psi_tdM_t+dL_t,\;\;Y_T=0.
\end{equation}
\ee{prop}

{\it Proof.} The variational principle gives that
$$E(H-X_T(\pi^*))X_T(\pi)=0,\;\;\forall\pi\in\Pi({\cal G}).$$

Since $\pi^*\in\Pi({\cal G})$ we have that
$E\big(\int_0^T\pi^*_u\lambda_ud\la M\ra_u\big)^2<\infty$ and by
the GKW decomposition
\begin{equation}\label{pil}
-\int_0^T\pi^*_u\lambda_ud\la M\ra_u=c+\int_0^T\psi_udM_u+N_u, \;\;\;\;\la M, N\ra=0,
\end{equation}
where $\psi\cdot M$ and $N$ are square integrable martingales.
Using the martingale property, it follows
from (\ref{pil}) that the triple $(Y,\psi,N)$, where
$$
Y_t=E\big(\int_t^T \pi^*_u\lambda_ud\la M\ra_u|{\cal F}_t)
$$
and $\psi, N$ are difined by (\ref{pil}),
satisfies the BSDE
\begin{equation}\label{bsd}
Y_t=Y_0+\int_0^t\pi^*_u\ld_ud\la
M\ra_u+\int_0^t\psi_udM_u+N_t,\;\;Y_T=0.
\end{equation}
Note that $Y_0=c=E\int_0^T\pi^*_u\lambda_ud\la M\ra_u$.

Therefore (taking  in mind decompositions (\ref{hhh2}),
(\ref{pil})) we have
\begin{align}\notag
E(H&-X_T(\pi^*)) X_T(\pi)\\\notag
& =E\left(-\int_0^T\pi^*_t\ld_td\la M\ra_t-\int_0^T\pi^*_tdM_t+H\right)\left(\int_0^T\pi_tdS_t\right)\\
\notag
&=E\left(Y_0+\int_0^T\psi_tdM_t+N_T-\int_0^T\pi^*_tdM_t+H\right)
\left(\int_0^T\pi_tdS_t\right)\\
\lbl{mp} &
=E\left(Y_0+N_T+\int_0^T\left(\psi_t-\pi^*_t\right)dM_t+H\right)
\left(\int_0^T\pi_t\ld_td
\la M\ra_t\right)\\
\lbl{br}& +E\left(Y_0+N_T+
\int_0^T\left(\psi_t-\pi^*_t\right)dM_t+c^H+\int_0^Th_tdM_t+L_T\right)\left(\int_0^T\pi_tdM_t\right)=0.
\end{align}

Using the formula of integration by parts in (\ref{mp}) and  properties of
mutual characteristics of
martingales in (\ref{br})we obtain the equality
\begin{align*}
& E\int_0^T\left(Y_0+L_t+\int_0^t\left(\psi_u-\pi^*_u\right)dM_t+H_t\right)
\pi_t\ld_td\la
M\ra_t\\
& \qquad +E\int_0^T\left(\psi_t+h_t-\pi^*_t\right)\pi_td\la M\ra_t=0.
\end{align*}
Inserting the solution $Y$ of  BSDE (\ref{bsd}) in the latter equality  gives
$$
 E\int_0^T\left(Y_0+H_t+Y_t-\int_0^t\ld_u\pi^*_ud\la
M\ra_u-\int_0^t\pi^*_udM_t\right)\pi_t\ld_td\la
M\ra_t
$$
$$
+E\int_0^T\left(\psi_t+h_t-\pi^*_t\right)\pi_td\la M\ra_t
$$
$$
=E\int_0^T\big((H_t+Y_t-X_t^*)\ld_t +\psi_t+h_t-\pi^*_t\big)\pi_td\la M\ra_t=0.
$$
By arbitrariness of $\pi\in\Pi({\cal G})$ we get
$$
E\left[\left(H_t+Y_t-\int_0^t\pi^*_udS_u\right)
\ld_t+\psi_t+h_t-\pi^*_t|{\cal G}_t\right]=0
$$
which gives (\ref{str}). \qed

\be{cor} Let conditions A), B) and C) be satisfied. Then
(\ref{str}),(\ref{bsde}) is equivalent to
the system of
Forward-Backward equations
\begin{equation}\label{fbe22}
d\wh X^*_t =\left(\widehat h_t+\frac{d\la \widehat M,m\ra_t}{d\la
M\ra_t}+\ld_t(\widehat H_t+\widehat Y_t-\widehat X_t^*)\right)d\wh S_t,\;\;\;\wh X^*_0=x\\
\end{equation}
\begin{equation}\label{fbe23}
d\widehat Y_t =\ld_t\left(\widehat h_t+\frac{d\la \widehat M,m\ra_t}{d\la
M\ra_t}+\ld_t(\widehat H_t+\widehat Y_t-\widehat X_t^*)\right)d\la
M\ra_t+dm_t,\;\;\widehat Y_T=0.
\end{equation}
\ee{cor}
{\it Proof.}
Since $\ld,\pi^*,\la M\ra$ are
$\cal G$-adapted,
$$
E(\lambda_tH_t+\lambda_tY_t-\lambda_tX_t^*|{\cal G}_t)
=\lambda_t(\widehat H_t+\widehat Y_t-\widehat X_t^*)
$$
and
$$
m_t=\int_0^t\psi_sdM_s+L_t
$$
is a $\cal G$-martingale. Therefore, by (\ref{kfg1})
$$
\widehat\psi_t=E\big(\frac{d{\la M,m\ra_t}}{d\la
M\ra_t}|{{\cal G}_t}\big)=\frac{d\la \widehat M,m\ra_t}{d\la M\ra_t}
$$
and it follows from  (\ref{str}) and (\ref{bsde})   that the optimal strategy
$\pi^*$ satisfies the system
\begin{align}\lbl{pi*}
\pi^*_t& =\widehat h_t+\frac{d\la \widehat M,m\ra_t}{d\la
M\ra_t}+\ld_t(\widehat H_t+\widehat Y_t-\widehat X_t^*)\\
\lbl{why} d\widehat Y_t& =\pi^*_t\ld_td\la
M\ra_t+dm_t,\;\;\widehat Y_T=0.
\end{align}
If we insert the
expression (\ref{pi*}) for $\pi^*$ in (\ref{why}) and  then integrate  both parts of
equation  (\ref{pi*}) with respect to $\wh S$ we obtain
the system of Forward-Backward equations (\ref{fbe22}), (\ref{fbe23}). \qed

{\bf Remark 3.1.} If we use the GKW decomposition of $m$ with respect to $\wh M$
$$
m_t=\int_0^t\wt\psi_ud\wh M_u+\wt L_t,\;\;\;\;\la\wh M, \wt L\ra=0,
$$
then by  (\ref{kfg1}) $\wh\psi_t=\rho^2_t\wt\psi_t$ and one can write the Forward-Backward equations
(\ref{fbe22}), (\ref{fbe23}) in the form
$$
d\wh X^*_t=\left(\widehat h_t+\rho^2_t\wt\psi_t+\ld_t(\widehat H_t+\widehat Y_t-\widehat X_t^*)
\right)d\wh S_t,\;\;\;\wh X^*_0=x
$$
$$
d\widehat Y_t =\ld_t\left(\widehat h_t+\rho^2_t\wt\psi_t+\ld_t(\widehat H_t+\widehat Y_t-\widehat X_t^*)\right)d\la
M\ra_t+\wt\psi_td\wh M_t+d\wt L_t,\;\;\widehat Y_T=0.
$$

From now on we assume

$\bf E$) $\rho^2_t< 1$ for all $t\in[0,T]$.

Let us introduce the operator $AY$ defined for any ${\cal M}^2({\cal G},P)$ by
$$(AY)_t=E\left(\int_0^T\frac{1}{1-\frac{d\la \widehat M\ra_t}{d\la
M\ra_u}}\left[Y_u\ld_u+\frac{d\la \widehat M,Y\ra_t}{d\la
M\ra_u}\right]\left(\ld_ud\la M\ra_u+d\widehat M_u\right)\bigg
|{\cal G}_t\right).$$

We shall use the following notations;
$$\widetilde h_t=\widehat h_t -\frac{d\la \widehat M,\widehat
H\ra_t}{d\la M\ra_t},\;\;\wt H=\widehat H_T-\int_0^T\frac{\wt
h_t}{1-\frac{d\la \widehat M\ra_t}{d\la M\ra_t}}d\widehat S_t$$
Let us consider  equation
\begin{align}\lbl{meH1}
\widetilde Y_T=\wt
H-\int_0^T\frac{1}{1-\rho_t^2}\left[\ld_t\widetilde Y_t
+\rho_t^2\widetilde\psi_t \right]\left(\ld_td\la M\ra_t+d\widehat
M_t\right),
\end{align}
which can be written in the form $\wt Y_T=\wt H-(A\wt Y)_T$.

 \be{thr}\label{pr2} Let $E\tilde H^2<\infty$. Then equation $(\ref{meH})$
admits a unique solution $\widetilde Y\in {\cal M}^2({\cal G},P)$
satisfying $E|\widetilde Y_T|^2\le E|\wt H|^2.$ \ee{thr}

{\it Proof}.   We need only to show that $A$ is a non-negative
operator. Indeed, for $Y_t=c+\int_0^t\vp_sd\widehat M_s+L_t,\;\la
\widehat M,L\ra=0$ we have
\begin{align*}
& \left(Y,AY\right)\\
& =E\bigg(Y_T \int_0^T\frac{1}{1-\rho_t^2}Y_t\ld_t^2d\la
M\ra_t+Y_T\int_0^T\frac{1}{1-\rho_t^2} Y_t\ld_td\widehat M_t\\
& \quad + Y_T\int_0^T\frac{\rho_t^2}{1-\rho_t^2}\vp_t\ld_td\la
M\ra_t+Y_T\int_0^T\frac{\rho_t^2}{1-\rho_t^2}\vp_t d\widehat
M_t\bigg)
\end{align*}
Since $\la Y,\widehat M\ra_t=\int_0^t\varphi_u\rho^2_ud\la M\ra_u$
and $EY_T\int_0^T{g}_ud\la M\ra_u=E\int_0^TY_u{g}_ud\la M\ra_u$
for any ${\cal G}$-predictable process $g$, we obtain that
\begin{align*}
& \left(Y,AY\right)\\
 & =E\bigg( \int_0^T\frac{1}{1-\rho_t^2}Y_t^2\ld_t^2d\la
M\ra_t+\int_0^T\frac{1}{1-\rho_t^2}Y_t\ld_t\vp_td\la\widehat
M\ra_t\\
& \quad +\int_0^T\frac{\rho_t^2}{1-\rho_t^2}Y_t\vp_t\ld_td\la
M\ra_t+ \int_0^T\frac{\rho_t^2}{1-\rho_t^2}\vp_t^2d\la\widehat
M\ra_t \bigg)
\\
& =E\bigg( \int_0^T\frac{1}{1-\rho_t^2}Y_t^2\ld_t^2d\la
M\ra_t+\int_0^T\frac{\rho_t^2}{1-\rho_t^2}Y_t\ld_t\vp_td\la
M\ra_t\\
& \quad +\int_0^T\frac{\rho_t^2}{1-\rho_t^2}Y_t\vp_t\ld_td\la
M\ra_t+ \int_0^T\frac{\rho_t^4}{1-\rho_t^2}\vp_t^2d\la M\ra_t
\bigg)\\
&
=E\int_0^T\frac{1}{1-\rho_t^2}\left(Y_t\ld_t+\rho_t^2\vp_t\right)^2d\la
M\ra_ t\ge 0.
\end{align*}
 Thus $Y+AY$ is a strictly positive operator, $(Id+A)^{-1}$ is
bounded with the norm less than one and $Y=(Id+A)^{-1}\wt H$ is a
unique solution of (\ref{meH1}).\qed

{\bf Remark 3.2.} Condition   $E\tilde H^2<\infty$ is satisfied if $EH^2<\infty$ and
$\rho^2_t\le 1-\varepsilon$ for all $t\in[0,T]$, for some $\varepsilon >0$.

{\bf Remark 3.3.} If $(\tilde Y,\tilde\psi)$ is a solution of equation  (\ref{meH1}),then
it follows from the proof of Theorem 3.1 that
 \begin{equation}\label{vars}
 E\int_0^T\frac{1}{1-\rho_t^2}\left(\tilde Y_t\ld_t+\rho_t^2\tilde\vp_t\right)^2d\la
M\ra_ t= (\tilde Y, A\tilde Y)= E\tilde Y_T\tilde H-E\tilde Y_T^2<\infty.
\end{equation}

 \be{thr}\label{pr2} Let
conditions A)-E) be satisfied and let $E\tilde H^2<\infty$. Then the strategy $\pi^*$ is
optimal if and only if  it admits the representation
\beq\lbl{oppi}\pi^*_t=\frac{1}{1-\rho_t^2}\left(\widetilde h_t
+\ld_t\widetilde Y_t+{\rho_t^2\widetilde\psi_t}\right) \eeq
 where the
pair $(\tilde Y,\tilde\psi)$ satisfies equation $(\ref{meH1})$.
\ee{thr} {\it Proof.}
 Let us show that if  the strategy $\pi^*$ is optimal, then it is of the form (\ref{oppi}).
 By Proposition \ref{p2.2} $\widehat
X_t(\pi^*)=\int_0^t\pi^*_s\ld_sd\la
M\ra_s+\int_0^t\pi^*_sd\widehat M_s$. Introducing notations
$$\widetilde Y_t=\widehat Y_t+\widehat H_t-\widehat X_t(\pi^*),\;\;\widetilde m_t=m_t+\widehat
H_t-\int_0^t\pi^*_sd\widehat M_s$$
(note that $\tilde Y=\tilde m$ by (\ref{why})) we have
\begin{align*}
\pi^*_t& =\widehat h_t+\frac{d\la \widehat M,\widetilde
m\ra_t}{d\la M\ra_t}+\pi^*_t\frac{d\la \widehat M\ra_t}{d\la
M\ra_t}-
\frac{d\la \widehat M,\widehat H\ra_t}{d\la M\ra_t}+\ld_t\widetilde Y_t,\\
d\widetilde Y_t& =d\widetilde m_t,\;\;\widetilde Y_T=\widehat
H_T-\widehat X_T(\pi^*),
\end{align*}
which gives (since $\rho^2_t< 1$ for all $t$)
\begin{align}
\lbl{str1} \pi^*_t& =\frac{1}{1-{\rho_t^2}} \left[\widehat h_t- \frac{d\la \widehat M,\widehat
H\ra_t}{d\la
M\ra_t}+\frac{d\la \widehat M,\widetilde Y\ra_t}{d\la M\ra_t}+\ld_t\widetilde Y_t\right],\\
\lbl{bo} \widetilde Y_T& =\widehat H_T-\widehat X_T(\pi^*).
\end{align}
Integrating (\ref{str1}) with respect to $\widehat S$
$$
 \widehat X_T(\pi^*)=\int_0^T\frac{1}{1-\rho_t^2} \left[\widehat h_t- \frac{d\la \widehat M,\widehat
H\ra_t}{d\la
M\ra_t}+\frac{d\la \widehat M,\widetilde Y\ra_t}{d\la M\ra_t}+\ld_t\widetilde Y_t\right]d\widehat S_t
$$
and inserting the latter equality into (\ref{bo}) we obtain the equation for the
martingale $\widetilde Y$
\begin{align}\lbl{meH}
\widetilde Y_T=\wt H-\int_0^T\frac{1}{1-\rho_t^2}\left[\frac{d\la \widehat M,\widetilde
Y\ra_t}{d\la M\ra_t}+\ld_t\widetilde Y_t\right]\left(\ld_td\la
M\ra_t+d\widehat M_t\right).
\end{align}
We remark that if $\widetilde Y_t=\widetilde
Y_0+\int_0^t\widetilde\psi_sd\widehat M_s+L_t^{\cal G}$ is the GKW
decomposition of $\widetilde Y$ then (\ref{meH}) can be rewritten
as (\ref{meH1}).

Let us show now that if the strategy $\pi^*$ is of the form
(\ref{oppi}),  then it is optimal.  Let first verify that $\pi^*\in\Pi({\cal G})$.
It follows from Theorem 3.1 that
$E\big(\int_0^T\pi_u^*d\wh S_u\big)^2<\infty$. Therefore,
$$
 E\big(\int_0^T\pi_u^*dS_u\big)^2 =E\big(\int_0^T\pi_u^*d\wh S_u+\int_0^T\pi_u^*d(\wh M_u-M_u)\big)^2\le
$$
$$
2E\big(\int_0^T\pi_u^*d\wh S_u\big)^2+2E \int_0^T(\pi_u^*)^2d\la\wh M-M\ra_u<\infty,
$$
since it follows from (\ref{duw}) and (\ref{vars}) that
$$
E \int_0^T(\pi_u^*)^2d\la\wh M-M\ra_u=E \int_0^T(\pi_u^*)^2d\la\wh M-M\ra^p_u =
$$
$$
=E \int_0^T(\pi_u^*)^2(1-\rho_u^2)d\la M\ra_u =
$$
$$
E\int_0^T\frac{\tilde h^2_u}{1-\rho_u^2}d\la M\ra_u+
E\int_0^T\frac{1}{1-\rho_u^2}(\lambda\tilde Y_u+\rho^2_u\tilde\psi_u)^2d\la M\ra_u<\infty
$$
Thus $E\big(\int_0^T\pi_u^*dS_u\big)^2<\infty$ and by Theorem 4.9 from  \cite{Ch} (see also
\cite{D-M-S-S-S})
$$
E\int_0^T(\pi_u^*)^2\la M\ra_u\le const\;  E\big(\int_0^T\pi_u^*dS_u\big)^2<\infty
$$
and $\pi^*\in\Pi({\cal G})$ by Remark 2.3.

By the variational principle it
is sufficient to show that
\begin{equation}\label{var0}
E\left(H-\int_0^T\pi^*_udS_u\right)\left(\int_0^T\pi_udS_u\right)=0,\;\;\forall\pi\in\Pi.
\end{equation}
From equation  (\ref{meH}) we have that
$$
-\int_0^T\pi^*_u\lambda_ud\la M\ra_u=\widetilde Y_T-\wh
H_T+\int_0^T\pi^*_ud\widehat M_u.
$$
Therefore
\begin{align*} & E\left(H-\int_0^T\pi^*_udS_u\right)\left(\int_0^T\pi_udS_u\right)\\
& =E\bigg(\wt Y_T+H-\wh H_T
+\int_0^T\pi^*_ud(\widehat M_u-M_u)\bigg)\\
& \quad \times \bigg(\int_0^T \pi_u\lambda_ud\la
M\ra_u+\int_0^T\pi_udM_u\bigg).
\end{align*}

Since $\lambda$ and $\la M\ra$ are ${\cal G}$-adapted
$$
E\left(H-\wh H_T\right)\left(\int_0^T \pi_u\lambda_ud\la M\ra_u\right)=0
$$
and  by Proposition \ref{p2.2}
$$
E(\int_0^T\pi^*_ud(\widehat M_u-M_u)\left(\int_0^T \pi_u\lambda_ud\la
M\ra_u\right)
$$
$$=
E\int_0^T \pi_u\lambda_ud\la
M\ra_uE\left(\int_0^T\pi^*_ud(\widehat M_u-M_u)|{\cal
G}_T\right)=0.
$$
Since $\widetilde Y$ is a martingale
\begin{equation}\label{y11}
E\widetilde Y_T\left(\int_0^T \pi_u\lambda_ud\la M\ra_u\right)=
 E\int_0^T \pi_u\lambda_u\widetilde Y_ud\la M\ra_u.
\end{equation}
Using the GKW decomposition for $\widetilde Y_t$ and relations
(\ref{duw}), (\ref{dul})
\begin{equation}\label{y22}
E\widetilde Y_T\int_0^T \pi_udM_u= E\int_0^T
\pi_u\widetilde\psi_ud\la M,\widehat M\ra_u= E\int_0^T
\pi_u\widetilde\psi_u\rho^2_ud\la M\ra_u.
\end{equation}
Using decompositions (\ref{hhh2}), (\ref{htg}) for $H$, projection
theorem and again relations (\ref{duw}), (\ref{dul})
\begin{align}
\notag
& E(H-\wh H_T)\left(\int_0^T\pi_udM_u\right)\\
\notag & =E\int_0^T\pi_uh_ud\la M\ra_u-E\int_0^T\pi_uh^{\cal
G}_ud\la M,\widehat
M\ra_u \\
\label{y33} & =E\int_0^T\pi_u\tilde h_ud\la M\ra_u.
\end{align}
Taking the sum of right-hand sides of (\ref{y11}), (\ref{y22}) and
(\ref{y33}) we obtain
\begin{align} \notag
&E\int_0^T \pi_u\lambda_u\widetilde Y_ud\la M\ra_u +  E\int_0^T
\pi_u\widetilde\psi_u\rho^2_ud\la M\ra_u \\
\notag
&\quad+ E\int_0^T\pi_u\widetilde h_ud\la M\ra_u \\
\notag  &=E\int_0^T \pi_u(\widetilde h_u+\lambda_u\tilde
Y_u+\widetilde\psi_u\rho^2_u)d\la M\ra_u
\\
\label{y44} &=E\int_0^T \pi_u\pi^*_u(1-\rho^2_u)d\la M\ra_u.
\end{align}
Finally,
\begin{align*}
& E\int_0^T\pi^*_ud(\widehat M_u-M_u)\int_0^T\pi_udM_u
\\
& = E\int_0^T\pi^*_u\pi_ud\la\widehat M, M\ra_u -
E\int_0^T\pi^*_u\pi_ud\la M\ra_u \\
& =-E\int_0^T \pi_u\pi^*_u(1-\rho^2_u)d\la M\ra_u,
\end{align*}
which, together with (\ref{y44}), implies that (\ref{var0})
is fulfilled and hence $\pi^*$ is optimal.\qed

\section{Relations to BSDEs for the value process}

\

In this section we express the solution of equation (\ref{meH}) in
terms of the value process of the problem (\ref{mvh}) and show that
equation (\ref{meH}) is equivalent to the BSDE derived in \cite{mtt}.

To this end
we consider  equation
\begin{equation}\lbl{consh}
\widetilde Y_T=\zeta-\int_\tau^T\frac{1}{1-\rho_t^2}\left(\ld_t\widetilde
Y_t+\rho_t^2\widetilde\psi_t\right)\,d\wh{S}_t
\end{equation}
for any stopping time $\tau\le T$. Similarly to Theorem 3.1 one can show that if $E\zeta^2<\infty$, then there exists
a unique solution $(\wt Y,\wt\psi)$  of (\ref{consh}), where $\wt Y$ is a square integrable martingale.

\be{lem}\lbl{l1.2} Let $(\wt Y^\tau,\wt\psi^\tau)$ and $(\wt Y,\wt\psi)$ be  solutions  of
equations
\begin{equation}\lbl{cons}
\widetilde Y_T=c-\int_0^T\frac{1}{1-\rho_t^2}\left(\ld_t\widetilde
Y_t+\rho_t^2\widetilde\psi_t\right)\,d\wh{S}_t
\end{equation}
and
\begin{equation}\label{fit}
\wt Y_T^\tau=1-\int_\tau^T \frac{1}{1-\rho_u^2} \,(\lb_u\wt
Y_u^\tau+\rho^2_u\widetilde\psi^\tau_u)d\widehat S_u
\end{equation}
respectively.
{\it Proof.}
Let
$$ \wt\pi_u=
\frac{1}{1-\rho_u^2} \,(\lb_u\wt Y_u+\rho_u^2\wt\psi_u), \;\;\;\;
\wt\pi_u^\tau= \frac{1}{1-\rho_u^2} \,(\lb_u\wt
Y_u^\tau+\rho_u^2\wt\psi_u^\tau).
$$
Then
\begin{align}\label{rel}
\notag \wt Y_t&=\wt Y_t^\tau(c-\int_0^\tau\wt\pi_ud\wh
S_u),\;\wt\psi_t=\wt\psi_t^\tau(c-\int_0^\tau\wt\pi_ud\wh S_u),\\
\wt\pi_t&=\wt\pi_t^\tau(c-\int_0^\tau\wt\pi_ud\wh S_u),\;\;
t\ge\tau.
\end{align}
\ee{lem} {\it Proof}. Multiplying both parts of equation (\ref{fit})
by $c-\int_0^\tau\wt\pi_ud\wh
S_u$   we get
\begin{align*}
&\wt Y_T^\tau(c-\int_0^\tau\wt\pi_ud\wh
S_u)\\
&=c-\int_0^\tau\wt\pi_ud\wh S_u-(c-\int_0^\tau\wt\pi_ud\wh S_u)\\
&\quad\times\int_\tau^T \frac{1}{1-\rho_u^2} \,(\lb_u\wt
Y_u^\tau+\rho_u^2\wt\psi_u^\tau)\,d\wh{S}_u
\end{align*}
Since $c-\int_0^\tau\wt\pi_ud\wh S_u$ is ${\cal
G}_\tau$-measurable using properties of stochastic integrals we
have
\begin{align*}
&\wt Y_T^\tau(c-\int_0^\tau\wt\pi_ud\wh S_u)=
c-\int_0^\tau\wt\pi_ud\wh S_u\\
&\quad-\int_\tau^T \frac{1}{1-\rho_u^2} \,\bigg(\lb_u\wt
Y_u^\tau(c-\int_0^\tau\wt\pi_ud\wh S_u)
+\rho_u^2\wt\psi_u^\tau(c-\int_0^\tau\wt\pi_ud\wh
S_u)\bigg)\,d\wh{S}_u.
\end{align*}
On the other hand
$$
\widetilde Y_T=c-\int_0^\tau\wt\pi_ud\widehat S_u-\int_\tau^T\frac{1}{1-\rho_t^2}
\left(\ld_t\widetilde
Y_t+\rho_t^2\widetilde\psi_t\right)\,d\wh{S}_t
$$
and relations (\ref{rel}) follow by the uniqueness of a solution of equation (\ref{consh})
with $\zeta=c-\int_0^\tau\wt\pi_ud\wh S_u$.\qed

Let us define the process
\begin{align*}
\wt V_t&=E\bigg[\left(1-\int_t^T \frac{1}{1-\rho_u^2} \,(\lb_u\wt
Y_u^t+\rho_u^2\wt\psi_u^t)\,d\wh{S}_u\right)^2 +
\int_t^T\frac{1}{1-\rho_u^2}(\lb_u\wt
Y_u^t+\rho_u^2\wt\psi_u^t)^2\,d\la M\ra_u|{\cal G}_t\bigg].
\end{align*}
\be{lem}\lbl{l1.3}
$\wt V_t>0,a.s.$ for all $t\in[0,T]$  and the process
$$
\wt V_t(c-\int_0^t\wt\pi_udS_u)^2+\int_0^t\wt\pi_u^2(1-\rho_u^2)d\la
M\ra_u
$$
 is a martingale. \ee{lem}

{\it Proof.} It is evident that $\wt V_t$ is non-negative. Let us
show that it is strictly positive. Assume that there exist
$t\in[0,T],B\in {\cal G}_t$ such that $P(B)>0$ and
\begin{align*}
E&\bigg[\left(1-\int_t^T \frac{1}{1-\rho_u^2} \,(\lb_u\wt
Y_u^t+\rho_u^2\wt\psi_u^t)\,d\wh{S}_u\right)^2\\
&+ \int_t^T\frac{1}{1-\rho_u^2}(\lb_u\wt
Y_u^t+\rho_u^2\wt\psi_u^t)^2\,d\la M\ra_u|{\cal G}_t\bigg]I_B=0.
\end{align*}
This implies that
\begin{equation}\label{ibs}
I_B-\int_t^TI_B\wt\pi_u^td\wh
S_u=0,
\end{equation}
\begin{equation}\label{ibm}
\int_t^TI_B\wt\pi_u^{t2}(1-\rho_u^2)d\la M\ra_u=0.
\end{equation}

Since $\rho_u<1$, it follows from (\ref{ibm}) that $\int_t^TI_B\wt\pi_u^td\wh S_u=0$.
Therefore, from (\ref{ibs}) we obtain $I_B=0$ a.s., which gives a
contradiction. Thus $P(B)=0$ and $\wt V$ is strictly positive.

Let us check now the martingale property. Using elementary
properties of conditional expectations and stochastic integrals it follows
from Lemma \ref{l1.2} that
\begin{align*}
& \wt V_t(c-\int_0^t\wt\pi_ud\wh S_u)^2\\
& =E\bigg[\left(c-\int_0^t\wt\pi_ud\wh S_u-(c-\int_0^t\wt\pi_ud\wh
S_u)\int_t^T\wt\pi_u^t\,d\wh{S}_u\right)^2\\
&\quad +(c-\int_0^t\wt\pi_ud\wh
S_u)^2\int_t^T(1-\rho_u^2)|\wt\pi_u^t|^2\,d\la M\ra_u|{\cal
G}_t\bigg]
\\
&=E\bigg[\left(c-\int_0^t\wt\pi_ud\wh S_u-\int_t^T
(c-\int_0^t\wt\pi_ud\wh S_u)
\wt\pi_u^t\,d\wh{S}_u\right)^2\\
&\quad+\int_t^T(1-\rho_u^2)|(c-\int_0^t\wt\pi_ud\wh
S_u)\wt\pi_u^t|^2\,d\la M\ra_u|{\cal G}_t\bigg]
\\
&=E\bigg[\left(c-\int_0^t\wt\pi_ud\wh S_u-\int_t^T
\wt\pi_ud\wh{S}_u\right)^2 +\int_t^T(1-\rho_u^2)\wt\pi_u^2d\la
M\ra_u|{\cal G}_t\bigg].
\end{align*}
Therefore, for any $t\in[0,T]$
$$
\wt V_t(c-\int_0^t\wt\pi_udS_u)^2+\int_0^t\wt\pi_u^2(1-\rho_u^2)d\la M\ra_u=
$$
$$
+E\bigg[\left(c-\int_0^T \wt\pi_ud\wh{S}_u\right)^2
+\int_0^T(1-\rho_u^2)\wt\pi_u^2d\la M\ra_u|{\cal G}_t\bigg],
$$
which proves that this process is a martingale.

\be{prop}\lbl{p3.4} The solution of (\ref{cons}) is strictly positive, i.e.,
$\wt Y_t>0$ a.s. for all $t\in[0,T]$.
 \ee{prop} {\it Proof.}
Let first show that $E\wt Y_T>0$. Multiplying both parts of equation (4.2) by $\wt Y_T$ and
taking expectations (as in the proof of Theorem 3.1) we obtain that
$$
E\wt Y_T^2=cE\wt Y_T-\int_0^T\frac{1}{1-\rho_u^2}(\wt Y_u\lambda_u+\rho_u^2\wt\psi_u)^2d\la M\ra_u.
$$
Therefore $cE\wt Y_T\ge E\wt Y_T^2>0$, hence $E\wt Y_T>0$.

Let us consider the process \beq\lbl{zz} Z_t=\wt
Y_t(c-\int_0^t\wt\pi_ud\wh S_u)+\int_0^t\wt\pi_u^2(1-\rho_u^2)d\la
M\ra_u. \eeq It follows from the Ito formula that $Z$ is a
martingale and using the martingale property from (\ref{cons}) we
have \beq\lbl{z4} \wt Y_t(c-\int_0^t\wt\pi_ud\wh S_u)=E(\wt
Y_T^2+\int_t^T\wt\pi_u^2(1-\rho_u^2)d\la M\ra_u|{\cal G}_t). \eeq
Besides the process $\wt Z_t=\wt Y_t(c-\int_0^t\wt\pi_ud\wh S_u)$
is a supermartingale and

\beq\lbl{z5} \wt Y_t(c-\int_0^t\wt\pi_ud\wh S_u)\ge E(\wt
Y_T^2|{\cal G}_t). \eeq

Let us define $\tau=\inf\{t:\wt Y_t=0\}\wedge T$ .  Then $\tau$ is a predictable stopping
time and there exists a  sequence of stopping
times $(\tau_n;n\ge1)$ such that $\lim\tau_n=\tau$ and
$\tau_n<\tau$ for every $n$ on $\tau>0$.
Note that $\wt Y_{\tau_n}>0$ by definition of $\tau_n$, since
$\wt Y_0=E\wt Y_T>0$.

Taking $\tau_n$ instead of $t$ in (\ref{z5}) and dividing both parts of this inequality by
$\wt Y_{\tau_n}$, we obtain

\begin{equation}\label{ec}
E\left(\frac{\wt Y_T^2}{\wt Y_{\tau_n}^2}\big|{\cal
G}_{\tau_n}\right)\le\frac{c-\int_0^{\tau_n}\wt\pi_ud\wh S_u}{\wt
Y_{\tau_n}}.
\end{equation}
It follows from the Lemma \ref{polk} (applied for the martingale
$\wt Y_t=E(\wt Y_T|{\cal G}_t)$) that
\begin{equation}\label{cont}
E\left(\frac{\wt Y_T^2}{\wt Y_{\tau_n}^2}\big|{\cal
G}_{\tau_n}\right)\to\infty\;\text{on the set}\;\{\wt Y_\tau=0\}.
\end{equation}

By Lemma \ref{l1.3} and (\ref{z4}) the processes
$\wt V_t(c-\int_0^t\wt\pi_ud\wh S_u)^2+\int_0^t\wt\pi_u^2(1-\rho_u^2)d\la
M\ra_u$ and $\wt
Y_t(c-\int_0^t\wt\pi_ud\wh S_u)+\int_0^t\wt\pi_u^2(1-\rho_u^2)d\la
M\ra_u$ are martingales and their values at time $T$ coincide, hence
they are undistinguishable. Thus
\begin{equation}\label{vy}
\wt V_t(c-\int_0^t\wt\pi_ud\wh S_u)^2=\wt
Y_t(c-\int_0^t\wt\pi_ud\wh S_u)
\end{equation}
which, together with (\ref{ec}), implies that
$$
E\left(\frac{\wt Y_T^2}{\wt Y_{\tau_n}^2}\big|{\cal
G}_{\tau_n}\right)\le\frac{c-\int_0^{\tau_n}\wt\pi_ud\wh S_u}{\wt
Y_{\tau_n}}=\frac{1}{\wt V_{\tau_n}}.
$$
Since $\wt V_t>0$, it follows  from the latter inequality
 $$
\lim_{n\to\infty}E\left(\frac{\wt Y_T^2}{\wt
Y_{\tau_n}^2}\big|{\cal G}_{\tau_n}\right)<\infty\;\text{on the
set}\;\{\wt Y_\tau=0\}
$$
which   contradicts to (\ref{cont}). Therefore $P(\wt
Y_\tau=0)=0$ and hence $\wt Y_t>0$ for all $t\in[0,T]$.

\be{cor} For all $t\in[0,T]$
\begin{equation}\label{yyy}
c-\int_0^t\wt\pi_ud\wh S_u\ge\wt Y_t
\end{equation}
  and
\begin{equation}\label{yyyy}
\wt V_t=\frac{\wt Y_t}{c-\int_0^t\wt\pi_ud\wh S_u}.
\end{equation}
\ee{cor}
{\it Proof.} By (\ref{z5}) and the Jensen inequality
\beq
 \wt Y_t(c-\int_0^t\wt\pi_ud\wh S_u)\ge E(\wt
Y_T^2|{\cal G}_t)\ge\wt Y^2_t. \eeq Since $\wt Y_t>0$ we obtain
inequality (\ref{yyy}). Therefore the process
$c-\int_0^t\wt\pi_ud\wh S_u$ is also strictly positive and
equality (\ref{yyyy}) follows from (\ref{vy}).

{\bf Remark 4.1.} $\wt V_t$ coincides with the value process $V_t$ of optimization problem
$$\underset{{\pi\in\Pi({\cal G})}}\min E(1-\int_0^T\pi_udS_u)^2$$ defined by
$$V_t=\underset{{\pi\in\Pi({\cal G})}}\essinf E((1-\int_t^T\pi_udS_u)^2|{\cal G}_t).$$
This follows from Theorem 3.2 and from Theorem 3.1 of \cite{mtt}.   But we shall show this equality, proving that
$\wt V$ satisfies the BSDE  for the value process $V$, derived in \cite{mtt}.

\begin{prop}\lbl{pro1.5}
Let $(\wt Y_t,\wt\psi_t)$ satisfies the equation
\begin{equation}\label{pl1}
    \wt Y_T=c-\int_0^T \frac{1}{1-\rho_t^2} \,(\lb_t\wt Y_t+\rho_t^2\wt\psi_t)\,d\wh{S}_t
\end{equation}
and let $\pi_t^*=\dfrac{1}{1-\rho_t^2}\,(\lb_t\wt
Y_t+\rho_t^2\wt\psi_t).$ Then $c-(\pi^* \cdot \wh{S})_t\equiv
c-\wh X_t^{\pi^*}$ is strictly positive and
\begin{equation}\label{ut}
U_t=\dfrac{\wt Y_t}{c-\wh
X_t^{\pi^*}}
\end{equation}
 is a solution of  BSDE
\begin{equation}\label{pl2}
    dU_t=\frac{(\lb_tU_t+\rho_t^2\psi_t)^2}{1-\rho_t^2+\rho_t^2 U_t}\,d\langle M\rangle_t+
        \psi_t\,d\wh{M}_t+dL_t, \quad U_T=1.
\end{equation}
\end{prop}

\begin{proof}
By Corollary 4.1 and Lemma \ref{p3.4}  $c-\wh X_t^{\pi^*}>0$
$P$-a.s. for all $t$. Therefore $U_t$ is a $\cal
G$-semimartingale. This semimartingale admits the decomposition
$$
U_t=A_t+\int_0^t\psi_sd\wh M_s+L_t,
$$
where $A_t$ is $\cal G$-predictable process of finite variation
and $L$ is a $\cal G$-local martingale strongly orthogonal to $\wh
M$.

By the It\^o  formula
$$
d\wt Y_t  =d( (c-\wh X_t^{\pi^*})U_t)
$$
$$
=(c-\wh X_t^{\pi^*})(dA_t+\psi_td\wh{M}_t+dL_t)
     -U_t \pi_t^*d\wh{S}_t-\pi_t^*\psi_t\rho_t^2 d\langle M\rangle_t
$$
$$
=( (c-\wh X_t^{\pi^*})\psi_t-\pi_t^*U_t)d\wh{M}_t
    +(c-\wh X_t^{\pi^*})dL_t
$$
\begin{equation}\label{diad}
     +(c-X_t^{\pi^*})dA_t-(\lb_tU_t\pi_t^*
    +\rho_t^2\psi_t\pi_t^*)d\langle M\rangle_t.
\end{equation}

Since $\wt Y$ is a martingale with the decomposition
\begin{equation}\label{comp}
\wt Y_t=\wt Y_0+\int_0^t\wt\psi_ud\wh M_u+\wt L_t
\end{equation}
comparing the decomposition terms of (\ref{diad}) and (\ref{comp})  we have

\begin{equation}\lbl{q123}
 \wt\psi_t=(c-\wh X_t^{\pi^*})\psi_t-\pi_t^* U_t,
\end{equation}
\begin{equation}\label{atu}
    A_t=\int_0^t\frac{\lb_tU_s+\rho_s^2\psi_s}{c-\wh X_s^{\pi^*}}\,\pi_s^*d\la M\ra_s.
\end{equation}
From (\ref{ut}) and (\ref{q123})
$$
 \pi_t^*=\dfrac{1}{1-\rho_t^2}\,(\lb_t\wt
Y_t+\rho_t^2\wt\psi_t)=
$$
$$
=\frac{1}{1-\rho_t^2}\,\left( \lb_tU_t(c-\wh X_t^{\pi^*})-
        \rho_t^2 U_t\pi_t^*-\rho_t^2(c-\wh X_t^{\pi^*})\psi_t\right)
$$
which gives
\begin{equation}\label{bol}
    \pi_t^*=\frac{\lb_tU_t+\rho_t^2\psi_t}{1-\rho_t^2+\rho_t^2U_t}\,
        (c-\wh X_t^{\pi^*}).
\end{equation}

Finally from (\ref{bol}) and (\ref{atu}) we obtain the equality
$$
A_t=\int_0^t\dfrac{(\lb_sU_s+\rho^2_s\psi_s)^2}{1-\rho_s^2+\rho_s^2U_s}d\la M\ra_s
$$
which means that $U_t$ satisfies (\ref{pl2}).
\end{proof}

\be{prop}\lbl{pro1.4} Let the triple $(V_t^{H},V_t,\wh X_t^{\pi^*})$
satisfy the Forward-Backward Stochastic differential equation
\begin{align}\label{trpl1}
    dV_t & =\frac{(\lb_tV_t+\rho_t^2\vp_t)^2}{1-\rho_t^2+\rho_t^2V_t}\,d\langle M\rangle_t +
        \vp_t\,d\wh{M}_t+dL_t, \quad V_T=1, \\
\label{trpl2}
    dV_t^{H} & =\frac{(\lb_tV_t+\rho_t^2\vp_t)(\lb_tV_t^{H}+\rho_t^2\vp_t^H)}
            {1-\rho_t^2+\rho_t^2V_t}\,d\langle M\rangle_t \notag \\
    & \quad + \vp_t^Hd\wh{M}_t+dL_t^H, \quad V_T^{H}=\wt H, \\
\label{trpl3}
    \pi_t^* & =\frac{\lb_tV_t^{H}+\rho_t^2\vp_t^H-\wh X_t^{\pi^*}(\lb_tV_t+\rho_t^2\vp_t)}
            {1-\rho_t^2+\rho_t^2V_t}\,, \quad \wh X_0^{\pi^*}=0, \\
    & \langle L^H,\wh{M}\rangle=\langle L,\wh{M}\rangle=0.  \notag
\end{align}

Then the pair $(\wt Y, \wt\psi)$, where
\begin{equation}\label{pair}
\wt Y_t=V_t^{H}-\wh X_t^{\pi^*}V_t\;\;\;\text{and}\;\;\;\;\
\wt\psi_t=\vp_t^H-V_t\pi^*_t-\vp_t\wh X_t^{\pi^*},
\end{equation}
is a solution  of equation
\begin{equation}\label{ple}
    \wt Y_T=\wt H-\int_0^T \frac{1}{1-\rho_t^2}\,(\lb_t\wt Y_t+\rho_t^2\wt\psi_t)\,d\wh{S}_t.
\end{equation}
\ee{prop}

\begin{proof}
By the It\^o formula
\begin{align*}
    d\wt Y_t=\,& -(V_t\pi^*_t+\wh X_t^{\pi^*}\vp_t-\vp_t^H)\,d\wh{M}_t
    +\wh X_t^{\pi^*}dL_t-dL_t^H \\
    & -\bigg( V_t\lb_t\pi^*_t+\wh X_t^{\pi^*}\,
        \frac{(\lb_tV_t+\rho_t^2\vp_t)^2}{1-\rho_t^2+\rho_t^2V_t}
     +\rho_t^2\vp_t\pi^*_t\\
     & -\frac{(\lb_tV_t^{H}+\rho_t^2\vp_t^H) (\lb_tV_t+\rho_t^2\vp_t)}{1-\rho_t^2+\rho_t^2V_t}
        \bigg)\,d\langle M\rangle_t.
\end{align*}
It follows from (\ref{trpl3}) that  the expression in the latter bracket is equal to
zero. Thus $\wt Y_t$ is martingale and
$\wt\psi_t=\vp_t^H-V_t\pi^*_t-\psi_t\wh X_t^{\pi^*}$. By (\ref{pair})
$$
\wt Y_T=\wt H-\wh X_T^{\pi^*}
$$
and inserting $(\wt Y,\wt\psi)$ in (\ref{ple}) we claim
$$
    \wh X_T^{\pi^*}
  =\int_0^T \frac{1}{1-\rho_t^2}\,(\lb_tV_t^{H}-\lb_tV_t\wh X_t^{\pi^*}
    +\rho_t^2\vp_t^H-\rho_t^2V_t\pi^*_t-\rho_t^2\wh X_t^{\pi^*})\,d\wh{S}_t.
$$
This means that
$$
\pi^*_t=\frac{1}{1-\rho_t^2}\, (\lb_tV_t^{H}+\rho_t^2\vp_t^H-\wh
X_t^{\pi^*}(\lb_tV_t+\rho_t^2\vp_t)- \rho_t^2V_t\pi^*_t)
$$
or
$$
    (1-\rho_t^2+\rho_t^2V_t)\pi^*_t=
    \lb_tV_t^{H}+\rho_t^2\vp_t^H-\wh X_t^{\pi^*}(\lb_tV_t+\rho_t^2\vp_t).
$$
Obviously this equality  coincides with (\ref{trpl3}). Therefore
$(\wt Y,\wt\psi)$ satisfies (\ref{ple}).
\end{proof}

\

\section{Diffusion market model}

\

Let us consider  the financial market model
$$
d\td S_t=\td S_t\mu_t(\eta)dt+\td S_t\sigma_t(\eta)dw_t^0,
$$
$$
d\eta_t=a_t(\eta)dt+b_t(\eta)dw_t,
$$
subjected to initial conditions, where only the second component
$\eta$ is observed. Here $w^0$ and $w$ are correlated Brownian
motions with $Edw^0_tdw_t=\rho dt, \rho\in(-1,1)$.

Let us write
$$
w_t=\rho w_t^0+\sqrt{1-\rho^2}w_t^1,
$$
where  $w^0$ and $w^1$ are independent Brownian motions.
It is evident that $w^\bot=-\sqrt{1-\rho^2}w^0+\rho w^1$ is a
Brownian motion independent of $w$  and one can express Brownian
motions $w^0, w^1$ in terms of $w$ and $w^\bot$ as
\begin{equation}\label{bm}
w_t^0=\rho w_t-\sqrt{1-\rho^2}w_t^\bot, \;\;
w_t^1=\sqrt{1-\rho^2}w_t+\rho w_t^\bot.
\end{equation}

We  assume that $b^2>0,\;\sigma^2>0$ and coefficients $\mu,
\sigma, a$ and $b$ are such that ${\cal F}_t^{S,\eta}={\cal
F}_t^{w^0,w}$, ${\cal F}^\eta_t={\cal F}_t^w$. So the stochastic
basis will be $(\Omega,{\cal F}, {\cal F}_t, P)$, where ${\cal
F}_t$ is the natural filtration of $(w^0, w)$ and the flow of
observable events is ${\cal G}_t=F^w_t$.

Also denote $dS_t=\mu_tdt+\sigma_tdw^0_t$, so that
$d\tilde S_t=\tilde S_tdS_t$
and $S$ is the return of the stock.

Let $\td\pi_t$ be the number shares of the stock
at time $t$. Then $\pi_t=\td\pi_t\td S_t$ represents an amount of money
invested in the stock at the time $t\in[0,T]$.
We consider the mean variance hedging problem
\begin{equation}\lbl{Imv}
\text{to minimize}\;\;\; E[ (x+\int_0^T\td\pi_td\td
S_t-H)^2]\;\;\;\;\text{over all}\;\;\td\pi\;\;\text{for
which}\;\;\td\pi\td S\in\Pi({\cal G}),
\end{equation}
which is equivalent to study the mean variance hedging problem
$$
\text{to minimize}\;\;\;\;\; E[ (x+\int_0^T\pi_td
S_t-H)^2]\;\;\;\;\text{over all}\;\;\;\;\pi\in\Pi({\cal G}).
$$
\begin{rem} Since $S$ is not ${\cal G}-$adapted,
$\wt\pi$ and $\wt\pi\wt S$ can not be simultaneously $\cal
G$-predictable
 and the problem
\begin{equation}\lbl{Imv1}
\text{to minimize}\;\;\;\;\; E[ (x+\int_0^T\td\pi_td\td
S_t-H)^2]\;\;\;\;\text{over all}\;\;\;\;\td\pi\in\Pi({\cal G}),
\end{equation}
is not equivalent to the problem (\ref{Imv}) and it needs separate consideration.
\end{rem}

Comparing with (\ref{str22}) we get that in this case
$$
M_t=\int_0^t\sigma_sdw_s^0,\;\;\;\langle M\rangle_t=\int_0^t\sigma_s^2ds,\;\;\;
\ld_t=\frac{\mu_t}{\sigma_t^2}.
$$
It is evident that $w$ is a Brownian motion also with respect to the
filtration $F^{w^0,w^1}$ and condition B) is satisfied. Therefore
by Proposition \ref{p2.2}
$$
\widehat
M_t=\rho\int_0^t\sigma_sdw_s.
$$
By the integral representation theorem the GKW decompositions
(\ref{hhh2}), (\ref{htg}) take following forms
\begin{equation}\label{hhh3}
c_H=EH,\;\;H_t=c_H+\int_0^t h_s\sg_sdw_s^0+\int_0^th_s^1 dw_s^1,
\end{equation}
\begin{equation}\label{htg3}
H_t=c_H+\rho\int_0^t h_s^{\cal G}\sigma_sdw_s+\int_0^th_s^\bot
dw_s^\bot.
\end{equation}
Putting expressions (\ref{bm}) for $w^0, w^1$ in (\ref{hhh3})
and equalizing integrands of (\ref{hhh3}) and (\ref{htg3}) we obtain that
$$
h_t=\rho^2h^{\cal G}_t-\sqrt{1-\rho^2}\frac{h_t^\bot}{\sigma_t}
$$
and hence
$$
\widehat h_t=\rho^2\widehat{h_t^{\cal G}}-\sqrt{1-\rho^2}\;
\frac{\widehat{h}_t^\bot}{\sigma_t}.
$$
Therefore by definition of $\widetilde h$
\begin{equation}\lbl{h}
\widetilde h_t=\rho^2 \widehat{h_t^{\cal G}}-\widehat h_t
=\sqrt{1-\rho^2}\;\frac{\widehat{h}_t^\bot}{\sg_t}.
\end{equation}

 We assume that $\sigma>0$. It is evident that $\frac{d\la \widehat M\ra_t}{d\la
M\ra_t}=\rho^2$ and (\ref{meH}) takes the form
\begin{equation}\label{meW}
\widetilde Y_T=\wt H-\frac{1}{1-\rho^2}\int_0^T\widetilde
Y_t\theta_t\left(\theta_tdt+\rho
dw_t\right)-\frac{\rho}{1-\rho^2}\int_0^T\widetilde\vp_t\left(\theta_tdt+\rho
dw_t\right)
\end{equation}
for $\widetilde Y_t=c+\int_0^t\widetilde\psi_s\rho\sigma_sdw_s
\equiv c +\int_0^t\widetilde\vp_sdw_s,$ where
$\theta_t=\frac{\mu_t}{\sg_t}$.
 We should solve (\ref{meW}) in the space of square integrable ${\cal G}$-martingales
$(\widetilde Y_t,{\cal G}_t)$. One can write this equation
with respect to random variable $\xi$
\begin{equation}\label{me}
\xi\!=\!\wt H-\frac{1}{1-\rho^2}\int_0^TE[\xi|{\cal
F}_t^w]\theta_t\left(\theta_tdt+\rho
dw_t\right)\!-\!\frac{\rho}{1-\rho^2}\int_0^TE[D_t\xi|{\cal
F}_t^w]\left(\theta_tdt+\rho dw_t\right),
\end{equation}
where $D$ is the stochastic derivative.

The optimal strategy in this case is
\beq\lbl{oppi*}\pi^*_t=\frac{1}{1-\rho^2}\left(\theta_t\widetilde Y_t+{\rho\widetilde\vp_t
+\sqrt{1-\rho^2}\wh h_t^\bot}\right)\sg_t^{-1} \eeq

\begin{rem} If $\rho=0$ and $\theta$ is deterministic then the
equation (\ref{me}) takes the form
\begin{equation}\label{me0}
\widetilde Y_T=\wt H-\int_0^T\widetilde Y_t\theta_t^2dt.
\end{equation} Performing the integration by part we obtain
$$\widetilde Y_T=c^H+\int_0^T\widetilde h_tdw_t-\widetilde Y_T\int_0^T\theta_s^2ds
+\int_0^T\int_0^t\theta_u^2dud\widetilde Y_t,$$
$$
\widetilde
Y_T\left(1+\int_0^T\theta_s^2ds\right)=c^H+\int_0^T\widetilde
h_tdw_t+\int_0^T\widetilde\vp_t\int_0^t\theta_u^2dudw_t.
$$
Hence
\begin{align*} \widetilde
Y_0=\frac{c^H}{1+\int_0^T\theta_s^2ds},\;\;\;
\widetilde\vp_t=\frac{\widetilde h_t}{1+\int_t^T\theta_s^2ds}.
\end{align*}
Therefore the solution of (\ref{me0}) can be given explicitly
\begin{align*} \widetilde
Y_t=\frac{c^H}{1+\int_0^T\theta_s^2ds}+\int_0^t\frac{\widetilde
h_s}{1+\int_s^T\theta_u^2du}dw_s
\end{align*}
and the optimal strategy is
\begin{align*}
\pi_t^*=\frac{c^H\ld_t}{1+\int_0^T\ld_s^2\sigma_s^2ds}+\ld_t\int_0^t\frac{\widetilde
h_s}{1+\int_s^T\ld_u^2\sigma_u^2du}dw_s.
\end{align*}
\end{rem}

\be{prop}\label{pr3} Suppose that $H=c^H,\;\; \eta_t=w_t$ and
$\frac{\mu_t}{\sigma_t}=\theta(t,w_t)$ for some continuous
function $\theta$, such that the nonlinear PDE
\begin{equation}\label{pde}
u_t +\frac{1}{2}u_{xx}=\frac{(\theta(t,x)u+\rho
u_x)^2}{1-\rho^2+\rho^2 u},\;\;\;\;u(T,x)=1
\end{equation} admits the sufficiently smooth solution $u$.
Then solution of $(\ref{meH})$ can be
represented as $$\widetilde Y_t=c^Hu(t,w_t){\cal
E}_{t}\left(-\int_0^\cdot\frac{\theta(s,w_s)u(s,w_s)+\rho
u_x(s,w_s)}{1-\rho^2+\rho^2 u(s,w_s)}(\theta(s,w_s)ds+\rho
dw_s)\right)$$ and optimal strategy is
\begin{align}\lbl{opttt}\notag
\pi^*_t&={c^H}{\sg^{-1}(t,w_t)}\frac{\theta(t,w_t)u(t,w_t)+\rho
u_x(t,w_t)}{1-\rho^2+\rho^2 u(t,w_t)}\\
&\times {\cal
E}_{t}\left(-\int_0^\cdot\frac{\theta(s,w_s)u(s,w_s)+\rho
u_x(s,w_s)}{1-\rho^2+\rho^2 u(s,w_s)}(\theta(s,w_s)ds+\rho
dw_s)\right)
\end{align}
 \ee{prop}

{\bf Sketch of the proof}. It is well known that the solution of
(\ref{pde}) defines by $V_t=u(t,w_t)$ the solution of
(\ref{trpl1}). On the other hand $V_t^H=cV_t$ and \\
$c-\wh
X_t^{\pi^*}=\mathcal{E}_t\bigg(-\int_0^\cdot
\frac{\lb_sV_s+\phi_s\rho_s^2}{1-\rho_s^2+\rho_s^2
V_s}\,d\wh{S}_s\bigg)$. Moreover, as in Proposition \ref{pro1.4}
it may be verified that $\wt Y_t=(c-\wh X_t^{\pi^*})V_t$ satisfies
equation (\ref{meH}). It follows from (\ref{bol}) that $\pi^*$ is of the form (\ref{opttt}).

The direct proof we shall give in the Appendix A.

{\bf Example}. If $\theta(t,x)=\theta(t)$ then the solution of
(\ref{pde}) is of the form $u(t,x)=u(t)$, where $u$ satisfies
$$\frac{du(t)}{dt}=\frac{\theta^2(t)u^2(t)}{1-\rho^2+\rho^2u(t)},\;\;\;u(T)=1.$$
Thus $$-\frac{1-\rho^2}{u(s)}+\rho^2\ln
u(s)\big|_t^T=\int_t^T\theta^2(s)ds.$$ Denote by
$\nu(\rho,\alpha)$ the unique roof of
$$\frac{1-\rho^2}{u}-\rho^2\ln u=\alpha$$ we can write
$u(t)=\nu(\rho,1-\rho^2+\int_t^T\theta^2(s)ds)$ and we get the solution of (\ref{me}) explicitly
$$\xi=c^H{\cal E}_T
\left(-\int_0^\cdot\frac{\theta(s)\nu\left(\rho,1-\rho^2+\int_s^T\theta^2(u)du\right)}
{1-\rho^2+\rho^2\nu\left(\rho,1-\rho^2+\int_s^T\theta^2(u)du\right)}\left(\theta(s)ds+\rho
dw_s\right)\right).$$ If in addition $\rho=0$ then
$\nu(0,y)=\frac{1}{y}$, $\xi=c^H\exp
\left(-\int_0^T\frac{\theta^2(t)}{1+\int_t^T\theta^2(s)ds}dt\right)
=\frac{c^H}{1+\int_0^T\theta^2(s)ds}$ and we obtain the solution
of equation (\ref{meW}) for deterministic $\theta$ and $H=c^H$
once again.

\

\appendix
\section{Appendix}

\

{\it The proof of Proposition \ref{pr3}}

It easy to see that (\ref{pde}) is equivalent to
\begin{equation*} u_t-\rho\frac{\theta(t,x)uu_x+\rho
u_x^2}{1-\rho^2+\rho^2 u}
+\frac{1}{2}u_{xx}=\frac{\theta^2(t,x)u^2+\rho
\theta(t,x)uu_x}{1-\rho^2+\rho^2 u},\;\;u(T,x)=1.
\end{equation*} Then if $u$ is a solution of (\ref{pde}) then using the
notation $g=-\frac{\theta(t,x)u+\rho u_x}{1-\rho^2+\rho^2 u}$, the
Feynmann-Kac formula and the Girsanov's Theorem we can write
$$u(t,x)=E\left({\cal E}_{tT}\left(\int_0^\cdot
g(s,w_s)(\theta(s,w_s)ds+\rho dw_s) \right)\big|w_t=x\right).$$
Hence the integrant $\td\vp$ for stochastic integral representation of the martingale \\
$\widetilde Y_t=c^HE\left[{\cal E}_T\left(\int_0^\cdot
g(s,w_s)(\theta(s,w_s)ds+\rho dw_s)\right)|{\cal F}_t^w\right]$
 can be calculated as follows
\begin{align*}
& \widetilde\vp_tdw_t=d\widetilde Y_t\\
& =c^Hd\left({\cal E}_{t}\left(\int_0^\cdot
g(s,w_s)(\theta(s,w_s)ds+\rho
dw_s)\right)u(t,w_t)\right)\\
& =c^H{\cal E}_{t}\left(\int_0^\cdot
g(s,w_s)(\theta(s,w_s)ds\!+\!\rho dw_s)\right)\\
& \quad \times\left(u_x(t,w_t)dw_t-g(t,w_t)
\left(\theta(t,w_t)u(t,w_t)\!+\!\rho u_x(t,w_t)\right)dt\right)\\
& \quad +c^Hu(t,w_t){\cal E}_{t}\left(\int_0^\cdot
g(s,w_s)(\theta(s,w_s)ds+\rho dw_s)\right)g(t,w_t)\left(\theta(t,w_t)dt+\rho dw_t\right)\\
& \quad +c^H{\cal E}_{t}\left(\int_0^\cdot
g(s,w_s)(\theta(s,w_s)ds+\rho
dw_s)\right)\rho g(t,w_t)u_x(t,w_t)dt\\
& =c^H{\cal E}_{t}\left(\int_0^\cdot g(s,w_s)(\theta(s,w_s)ds+\rho
dw_s)\right)(u_x(t,w_t)+\rho u(t,w_t)g(t,w_t))dw_t\\
\end{align*}
Thus
\begin{gather*}
\frac{1}{1-\rho^2}\int_0^T\widetilde
Y_t\theta(t,w_t)(\theta(t,w_t)dt+\rho
dw_t)+\frac{\rho}{1-\rho^2}\int_0^T\widetilde
\vp_t(\theta(t,w_t)dt+\rho dw_t) \\
=c^H\frac{1}{1-\rho^2}\int_0^T\left(u(t,w_t)\theta(t,w_t)+\rho
u_x(t,w_t)+\rho^2
g(t,w_t)u(t,w_t)\right) \\
\times{\cal E}_{t}\left(\int_0^\cdot g(s,w_s)(\theta(s,w_s)ds+\rho
dw_s)\right)(\theta(t,w_t)dt+\rho dw_t).
\end{gather*}
Since $u\theta+\rho u_x+\rho^2gu=(\rho^2-1)g$, then
\begin{gather*}
\frac{1}{1-\rho^2}\int_0^T\widetilde
Y_t\theta(t,w_t)(\theta(t,w_t)dt+\rho
dw_t)+\frac{\rho}{1-\rho^2}\int_0^T\widetilde
\vp_t(\theta(t,w_t)dt+\rho dw_t) \\
= -c^H\int_0^Tg(t,w_t){\cal E}_t\left(\int_0^\cdot
g(s,w_s)(\theta(s,w_s)ds+\rho
dw_s)\right)\left(\theta(t,w_t)dt+\rho dw_t\right).
\end{gather*}
On the other hand,
\begin{gather*}
\widetilde Y_T=c^H{\cal E}_T\left(\int_0^\cdot
g(s,w_s)(\theta(s,w_s)ds+\rho dw_s)\right) \\
=c^H+ \int_0^Tc^Hg(t,w_t){\cal E}_t\left(\int_0^\cdot g(s,w_s)
(\theta(s,w_s)ds+\rho dw_s)\right)(\theta(t,w_t)dt+\rho dw_t) .
\end{gather*}
Hence (\ref{meW}) is satisfied. The expression for $\pi^*$ is obtained from representation
\begin{align*}
\wt Y_T=&c_H{\cal
E}_{T}\left(-\int_0^\cdot\frac{\theta(s,w_s)u(s,w_s)+\rho
u_x(s,w_s)}{1-\rho^2+\rho^2 u(s,w_s)}(\theta(s,w_s)ds+\rho
dw_s)\right)
\\
=&-c_H\int_0^T\frac{\theta(s,w_s)u(s,w_s)+\rho
u_x(s,w_s)}{1-\rho^2+\rho^2 u(s,w_s)}
\\
&\times{\cal
E}_{s}\left(-\int_0^\cdot\frac{\theta u+\rho
u_x}{1-\rho^2+\rho^2 u}(\theta du+\rho
dw_u)\right)(\theta(s,w_s)ds+\rho
dw_s)
\end{align*}
and equations (\ref{meW})and (\ref{oppi*}).

\end{document}